\documentclass[preprint,showpacs,preprintnumbers,amsmath,amssymb,natbib]{revtex4}

\usepackage{graphicx}
\usepackage{epsfig}		
\usepackage{dcolumn}
\usepackage{bm}
\def\0{\mbox{\tiny $0$}}
\def\1{\mbox{\tiny $1$}}
\def\2{\mbox{\tiny $2$}}
\def\3{\mbox{\tiny $3$}}
\def\4{\mbox{\tiny $4$}}
\def\5{\mbox{\tiny $5$}}
\def\6{\mbox{\tiny $6$}}
\def\7{\mbox{\tiny $7$}}
\def\8{\mbox{\tiny $8$}}
\def\9{\mbox{\tiny $9$}}

\def\f14{\mbox{\tiny $\frac{1}{4}$}}

\def\N{\mbox{\tiny $N$}}
\def\R{\mbox{\tiny $R$}}
\def\U{\mbox{\tiny $U$}}
\def\F{\mbox{\tiny $F$}}

\def\s{\mbox{\tiny $s$}}

\def\mi{\mbox{\tiny $-$}}

\def\pl{\mbox{\tiny $+$}}
\def\al{\mbox{\tiny $\alpha$}}

\def\bb#1{\mbox{\footnotesize $(#1)$}}

\begin{document}

\preprint{DF/IST-01.2008}
\preprint{January 2008}

\title{Stationary condition in a perturbative approach for mass varying neutrinos}

\author{A. E. Bernardini}
\affiliation{Departamento de F\'{\i}sica, Universidade Federal de S\~ao Carlos, PO Box 676, 13565-905, S\~ao Carlos, SP, Brasil}
\email{alexeb@ifi.unicamp.br}
\author{O. Bertolami}
\affiliation{Instituto Superior T\'ecnico, Departamento de F\'{\i}sica, Av. Rovisco Pais, 1, 1049-001, Lisboa, Portugal}
\email{orfeu@cosmos.ist.utl.pt}
\altaffiliation[Also at]{~Instituto de Plasmas e Fus\~{a}o Nuclear}

\date{\today}

\begin{abstract}
A perturbative approach for arbitrary choices of the equation of state of the universe is introduced in order to treat scenarios for mass varying neutrinos (MaVaN's) coupled to the dark sector.
The generalized criterion for the applicability of such an approach is expressed through a constraint on the coefficient of the linear perturbation on the dark sector scalar field.
This coefficient depends on the ratio between the variation of the neutrino energy and the scalar field potential.
Upon certain conditions, the usual {\em stationary condition} found in the context of MaVaN models together with the perturbative contribution can be employed to predict the dynamical evolution of the neutrino mass.
Our results clearly indicate that the positiveness of the squared speed of sound of the coupled fluid and the model stability are not conditioned by the {\em stationary condition}.
\end{abstract}

\date{\today}

\maketitle

The ultimate nature of the dark energy and its connection with the Standard Model (SM) states is one of the most intriguing issues related with the negative pressure component required to understand the accelerated expansion of the universe.
If from one hand, the most natural candidate to couple a SM singlet quintessence-like scalar field is the Higgs sector \cite{Wil08,Ber08a}, an exciting alternative is the coupling to neutrinos as in the context of the
so-called mass varying neutrino (MaVaN) models \cite{Gu03,Far04,Pas06,Bja08,Ber08b}.
This last possibility is particularly interesting since the coupling of neutrinos to the dark energy scalar field component may lead to a number of significant phenomenological consequences.
Actually, this coupling renders the MaVaN mechanisms fairly natural.
Indeed, if the neutrino mass $m_{\nu}$ is generated by the dynamical value of a cosmologically active scalar field $\phi$ instead of through a vacuum expectation value (VEV) it would be an evolving quantity.

For the purpose of analyzing the minimal energy density contribution of neutrinos to the cosmic budget, a {\em stationary condition} (SC) has been devised \cite{Far04} so to establish a relationship between $m_{\nu}$ and the scalar field potential.
In this work we consider the possibility of neutrino masses arising from an interaction with the real scalar field that drives the accelerated expansion of the universe.
In order to obtain a model along these lines we consider the MaVaN contribution to the energy conservation equation of the cosmic fluid as a perturbation that iteratively modifies the background fluid equation of state.
This approach is taylored to satisfy the phenomenological requirement that the contribution of neutrinos to the energy density evolution is subdominant in the matter dominated era.
The scalar field is modified by a linear perturbation due to the neutrino mass coupling, which is {\em turned on} when the neutrinos become non-relativistic (NR).
In such a proposal (see Ref.~\cite{Ber08b} for a detailed discussion), one is compelled to abandon the SC even though this condition can be reconciliated with our perturbative approach through a particular neutrino mass dependence on the scale factor.

Another relevant related issue is that dark matter is most often not considered in the formulation of the MaVaN models.
However, the possibility of treating dark energy and dark matter in a unified scheme naturally offers this possibility.
The generalized Chaplygin gas (GCG) \cite{Kam02,Ber02} is particularly relevant in this respect as it is shown to be consistent with known observational constraints, from CMB \cite{Ber03}, supernova \cite{Ber04,Ber05}, gravitational lensing surveys \cite{Ber03B}, gamma ray bursts \cite{Ber06B} and cosmic topology \cite{Ber06}.
Our analysis considers the coupling of MaVaN's to the underlying scalar field of the generalized Chaplygin gas (GCG) model.
Since the neutrino contribution is perturbative, we obtain a small deviation from the stability condition characteristic of the unperturbed (GCG) equation of state.

Given that cosmological neutrinos have not been observed so far, hints at about their contribution to the energy density are relevant, albeit somewhat theoretical.
The neutrino energy density and pressure are expressed  through a Fermi-Dirac distribution function without a chemical potential, $f\bb{q}$, where $q \equiv \frac{|\mbox{\boldmath$p$}|}{T_{\nu\0}}$, $T_{\nu\0}$ being the neutrino background temperature at present.
Thus,
\begin{eqnarray}
\rho_{\nu}\bb{a, \phi} &=&\frac{T^{\4}_{\nu \0}}{\pi^{\2}\,a^{\4}}
\int_{_{0}}^{^{\infty}}{\hspace{-0.3cm}dq\,q^{\2}\,\left(q^{\2}+\frac{m^{\2}\bb{\phi}\,a^{\2}}{T^{\2}_{\nu\0}}\right)^{\1/\2}\hspace{-0.1cm}f\bb{q}},\\
p_{\nu}\bb{a, \phi} &=&\frac{T^{\4}_{\nu \0}}{3\pi^{\2}\,a^{\4}}\int_{_{0}}^{^{\infty}}{\hspace{-0.3cm}dq\,q^{\4}\,\left(q^{\2}+\frac{m^{\2}\bb{\phi}\,a^{\2}}{T^{\2}_{\nu \0}}\right)^{\mi\1/\2}\hspace{-0.1cm} f\bb{q}},~~~~ \nonumber
\label{gcg01}
\end{eqnarray}
where we have assumed a flat FRW cosmology, that the scale factor of the universe at present is $a_{\0} = 1$ (the sub-index $0$ denotes present-day values) and, for simplicity, we considered only one non-vanishing neutrino mass.
By observing that
\begin{equation}
m_{\nu}\bb{\phi} \frac{\partial \rho_{\nu}\bb{a, \phi}}{\partial m_{\nu}\bb{\phi}} = (\rho_{\nu}\bb{a, \phi} - 3 p_{\nu}\bb{a, \phi}),
\label{gcg02}
\end{equation}
and from Eq.~(\ref{gcg01}), one can obtain the energy-momentum conservation for the neutrino fluid
\begin{equation}
\dot{\rho}_{\nu}\bb{a, \phi} + 3 H (\rho_{\nu}\bb{a, \phi} + p_{\nu}\bb{a, \phi}) =
\dot{\phi}\frac{d m_{\nu}\bb{\phi}}{d \phi} \frac{\partial \rho_{\nu}\bb{a, \phi}}{\partial m_{\nu}\bb{\phi}},
\label{gcg03}
\end{equation}
where $H = \dot{a}/{a}$ is the expansion rate of the universe and the {\em overdot} denotes differentiation with respect to cosmic time ($^{\cdot}\, \equiv\, d/dt$).

It is important to emphasize that the coupling between cosmological neutrinos and the scalar field as specified in Eq.~(\ref{gcg02}) is restricted to times when neutrinos are NR, i. e. $\frac{\partial \rho_{\nu}\bb{a, \phi}}{\partial m_{\nu}\bb{\phi}} \simeq n_{\nu}\bb{a} \propto{a^{\mi\3}}$ \cite{Far04,Bja08,Pec05}.
On the other hand, as long as neutrinos are relativistic ($T_{\nu}\bb{a} = T_{\nu \0}/a >> m_{\nu}\bb{\phi\bb{a}}$), the decoupled fluids should evolve adiabatically since the strength of the coupling is suppressed by the relativistic increase of pressure ($\rho_{\nu}\sim 3 p_{\nu}$).
In this case, one should have
\begin{equation}
\dot{\rho}_{\phi} + 3 H (\rho_{\phi} + p_{\phi}) = 0,
\label{gcg04}
\end{equation}
and a similar equation for neutrinos.
For the scalar field background fluid one has
\begin{eqnarray}
\rho_{\phi} &=& \frac{\dot{\phi}^{\2}}{2} + V\bb{\phi},\nonumber\\
p_{\phi} &=& \frac{\dot{\phi}^{\2}}{2} - V\bb{\phi}.
\label{gcg05}
\end{eqnarray}

Treating the system of NR neutrinos and the scalar field as a unified fluid (UF) which adiabatically expands with energy density $\rho_{\U\F} = \rho_{\nu} + \rho_{\phi}$ and pressure $p_{\U\F} = p_{\nu} + p_{\phi}$ leads to
\begin{equation}
\dot{\rho}_{\U\F} + 3 H (\rho_{\U\F} + p_{\U\F}) = 0~~ \Rightarrow ~~\dot{\rho}_{\phi} + 3 H (\rho_{\phi} + p_{\phi}) = -\dot{\phi}\frac{d m_{\nu}}{d \phi} \frac{\partial \rho_{\nu}}{\partial m_{\nu}},
\label{gcg06}
\end{equation}
where the last step is derived from the substitution of Eq.~(\ref{gcg03}) in the first equation in (\ref{gcg06}).

It is well known that the relative contribution of the energy densities components of the universe with respect to the one of the dark energy sector is on its own a problem.
The assumptions proposed in Ref. \cite{Far04} and subsequently developed elsewhere \cite{Pec05,Bro06A,Bja08,Tak06} introduce a SC which allows circumventing the coincidence problem for cosmological neutrinos, by imposing that the dark energy is always diluted at the same rate as the neutrino fluid, that is,
\begin{equation}
\frac{d V\bb{\phi}}{d {\phi}} = - \frac{d m_{\nu}}{d \phi} \frac{\partial \rho_{\nu}}{\partial m_{\nu}}.
\label{gcg07}
\end{equation}
This condition introduces a constraint on the neutrino mass since it promotes it into a dynamical quantity, as indicated in Eq.~(\ref{gcg06}).
In this context, the main feature of the FNW scenario \cite{Far04} is that, in what concerns to dark sector, it is equivalent to a cosmological constant-like equation of state and an energy density that is as a function of the neutrino mass \cite{Ber08b}.
As already mentioned, the effectiveness of this coupling is limited by values of the scale factor greater than $a_{\N\R}$, where $a_{\N\R}$ parameterizes the transition between the relativistic and NR regimes.

The assumption of a universe with the dark energy governed by the equation of state $p_{\phi} = -\rho_{\phi}$ implies, through Eq.~(\ref{gcg07}), that  $\rho_{\Lambda} = V$ , and allows to recover the SC, exactly as obtained in Ref. \cite{Far04}.
It is clear that the relevance and the considerations about the constraints on the equation of state are conditioned by the underlying scalar field potentials and their adequacy concerning neutrino mass generation.
In fact, once one assumes that $p_{\phi} = -\rho_{\phi}$, the neutrino mass evolution and the form of the potential become directly entangled by the SC.
We believe that an alternative approach for treating deviations from the SC is needed since the constraint that it ensues, namely $\rho_{\phi} + p_{\phi} = 0$, is quite restrictive as it disregards the kinetic energy contribution of the scalar field. Although this
contribution is at present negligible in comparison to the one of the potential, it can dominate and play a role in the evolution of the universe at earlier times.

Given these considerations, instead of beginning from Eq.~(\ref{gcg06}), we use the unperturbed Eq.~(\ref{gcg04}), and establish the conditions for treating the neutrino coupling in a perturbative way.
To exemplify this suggestion more concretely, let us consider the unperturbed equation of motion for the scalar field
\begin{equation}
\ddot{\phi} + 3 H \dot{\phi} + \frac{d V\bb{\phi}}{d {\phi}} = 0,
\label{gcg08}
\end{equation}
and assume that the effect of the coupling of the neutrino fluid to the scalar field fluid is quantified by a linear perturbation $\epsilon \phi$ ($|\epsilon| << 1$) such that $\phi \rightarrow \varphi \approx (1 + \epsilon) \phi$. It then follows a novel equation for energy conservation
\begin{equation}
\ddot{\varphi} + 3 H \dot{\varphi} + \frac{d V\bb{\varphi}}{d {\varphi}} = -\frac{d m_{\nu}}{d \varphi} \frac{\partial \rho_{\nu}}{\partial m_{\nu}}.
\label{gcg10}
\end{equation}
After some straightforward manipulations one can easily find that
\begin{eqnarray}
\frac{d V\bb{\varphi}}{d {\varphi}}
&\simeq& \frac{d V\bb{\phi}}{d {\phi}} + (\epsilon \phi) \frac{d^{\2} V\bb{\phi}}{d {\phi^{\2}}},
\label{gcg11}
\end{eqnarray}
which, upon substitution into Eq.~(\ref{gcg10}), and use of Eq.~(\ref{gcg08}), leads to
\begin{eqnarray}
\epsilon \left[\phi \frac{d^{\2} V\bb{\phi}}{d {\phi^{\2}}} - \frac{d V\bb{\phi}}{d {\phi}} \right] \simeq \frac{d m_{\nu}}{d \varphi} \frac{\partial \rho_{\nu}}{\partial m_{\nu}} \simeq \frac{d m_{\nu}}{d \phi}
n_{\nu}\bb{a},
\label{gcg12}
\end{eqnarray}
where the perturbative character of the neutrino mass term is assumed when setting the last approximation in the above equation.
Finally, we obtain for the value of the coefficient of the perturbation
\begin{eqnarray}
\epsilon  \simeq \frac{-\frac{d m_{\nu}}{d \phi}\frac{\partial \rho_{\nu}}{\partial m_{\nu}}}{\left[\phi^{\2} \frac{d}{d {\phi}}\left(\frac{1}{\phi} \frac{d V\bb{\phi}}{d {\phi}}\right)\right]},
\label{gcg13}
\end{eqnarray}
which for consistency is required to satisfy the condition $|\epsilon| << 1$.
Upon fulfilling all known phenomenological requirements, the above result allows us to address a wide
class of scalar field potentials and related equations of state for various candidates for the dark sector
(dark energy and dark matter), which through the SC would be incompatible with realistic neutrino mass generation models.

Notice that the equation for the conservation of energy, Eq.~(\ref{gcg10}), can be reobtained by simply redefining $\rho_{\U\F}$ as
\begin{equation}
\rho_{\U\F}  = \frac{1}{2}\dot{\phi}^{\2} + V_{\mbox{\tiny Eff}}
\label{gcg14}
\end{equation}
so that the usual definition \cite{Bro06A,Far04,Bja08} of an {\em effective} potential $V_{\mbox{\tiny Eff}}$ in terms of $\frac{d V_{\mbox{\tiny Eff}}}{d \phi} = \frac{d V\bb{\phi}}{d \phi} + \frac{d m_{\nu}}{d \phi} \frac{\partial\rho_{\nu}}{\partial m_{\nu}}$ is now valid for any value of $\dot{\phi}$.

\begin{figure}
\vspace{-2.3 cm}
\centerline{\psfig{file=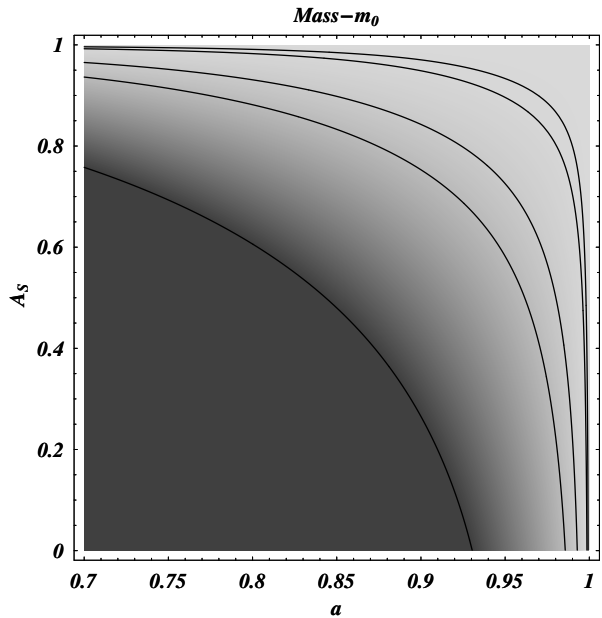,width=14cm}}
\vspace{-2.3 cm}
\caption{Present-day values of the neutrino mass $m_{\0}$ and the corresponding values of $a_{\N\R}$ for which the transition between the NR and UR regimes takes place in the GCG phenomenological scenario with $\alpha = 1/2$ and variable $A_{s}$.
The increasing {\em graylevel} corresponds to increasing values of $m_{\0}$, for which we have marked the boundary values for $m_{\0} = 0.05\, eV,\,0.1\, eV,\,0.5\, eV,\,1\, eV,\, 5\, eV$.}
\label{fGCG-03}
\end{figure}

\begin{figure}
\vspace{-2.3 cm}
\centerline{\psfig{file=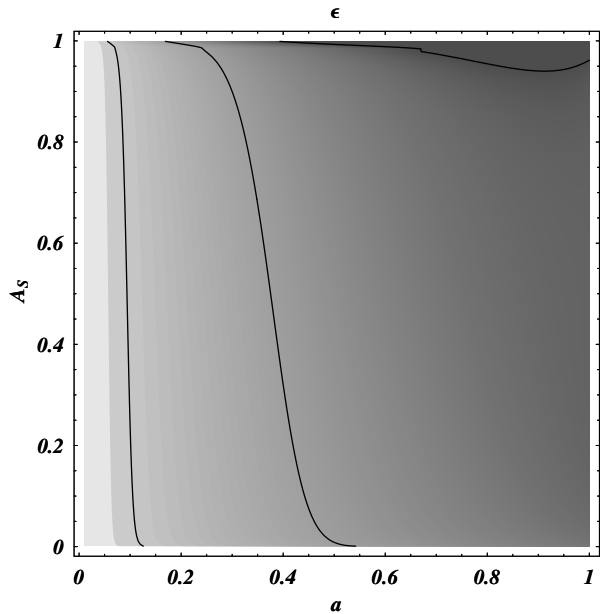,width=14cm}}
\vspace{-2.3 cm}
\caption{Maximal value of the linear perturbation coefficient $\epsilon$ as a function of the scale factor and of the GCG parameters $A_{\s} = 0.7$ and $\alpha = 1/2$.
For illustration purposes we have set $m_0 = 0.5\, eV$.
The increasing {\em gray level} corresponds to increasing values of $|\epsilon|$,
for which we have marked the boundary values for $|\epsilon|= 0.01, \,0.1, \,1$.}
\label{fGCG-04}
\end{figure}

In order to verify under which conditions Eq.~(\ref{gcg07}) agrees with our perturbative approach for a given background equation of state, the coefficient of the linear perturbation should be given by
\begin{eqnarray}
\epsilon  \simeq \frac{\frac{d V\bb{\phi}}{d \phi}}{\left[\phi^{\2} \frac{d}{d {\phi}}\left(\frac{1}{\phi} \frac{d V\bb{\phi}}{d {\phi}}\right)\right]}~~~~ (|\epsilon| << 1).
\label{gcg16}
\end{eqnarray}
This means that we must search for a neutrino mass dependence on the scale factor for which the above condition is satisfied.
Thus, once one sets the equation of state for the dark sector, there will be a period at late times for which the SC and the perturbative approach match.
In particular, this feature can be reproduced by the GCG equation of state given by,
\begin{equation}
p = - A_{\s} \rho_{\0} \left(\frac{\rho_{\0}}{\rho}\right)^{\al},
\label{gcg20}
\end{equation}
where $0 \leq A_{\s} \leq 1$ and $0 \leq \alpha \leq 1$ \cite{Kam02,Ber02}.
Inserting the above equation into the unperturbed energy conservation Eq. (\ref{gcg04}), one can integrate it to obtain \cite{Ber02}
\begin{equation}
\rho_{\phi} = \rho_{\0} \left[A_{\s} + \frac{(1-A_{\s})}{a^{\3(\1+\alpha)}}\right]^{\1/(\1 \pl \al)},
\label{gcg21}
\end{equation}
from which $p_{\phi}\bb{a}$ is easily computed, and by Eqs.~(\ref{gcg05}), the scalar potential $V\bb{\phi\bb{a}}$ \cite{Ber04} is written as
\begin{equation}
V\bb{\phi\bb{a}} = \frac{1}{2} \rho_{\0} \left[A_{\s} + \frac{(1-A_{\s})}{a^{\3(\1+\alpha)}}\right]^{-\al/(\1 \pl \al)}\left[2\,A_{\s} + \frac{(1-A_{\s})}{a^{\3(\1+\alpha)}}\right].
\label{gcg21B}
\end{equation}
Clearly, given a potential, the explicit dependence of $m\bb{\phi}$ on $a$ can be immediately obtained from Eq.~(\ref{gcg07}).
Furthermore, it is necessary to determine for which values of the scale factor the neutrino-scalar field coupling becomes important.
For convenience we set the value of $a = a_{\N\R}$ for which $\rho_{\nu,\N\R} = \rho_{\nu,\U\R}$ holds, usually established by the condition of $m_{\nu}\gtrsim T_{\nu}$, that parameterizes the transition between the NR and the ultra-relativistic (UR) regime.
In fact, this takes place when
\begin{equation}
m_{\nu}\bb{a} = \chi \frac{T_{\nu, \0}}{a},
\label{gcg33}
\end{equation}
where $\chi$ is a numerical factor estimated to be about $\chi \simeq 94$ considering that $\rho_{\nu}/\rho_{\mbox{\tiny Crit}} = m_{\0}\,[eV]/(94\,h^{\2}\,[eV])$, where $h$ is the value of the Hubble constant in terms of $100\, km\, s^{\mi\1}\,Mpc^{\mi\1}$.
Such a correspondence between $a_{\N\R}$ and $m_{\0}$ is illustrated in the Fig.~\ref{fGCG-03} for the particular value of $\alpha = 1/2$.
Considering the whole set of parameters that characterize the background fluid, one notices that is rather difficult to see that the maximal value assumed by the $\epsilon$ parameter corresponds to its present-day value, as seen in the Fig.~\ref{fGCG-04}.

For a more accurate comparison between our perturbative approach and the SC framework, it would be interesting to express analytically the bound ($\epsilon < 1$) for the coupling function in terms of the model parameters.
Considering dark sector scalar fields and potentials with a simple analytical dependence on the scale factor, we can
compare the constraint imposed on the potential by the breaking of the perturbative approach with the known limits
derived in the literature for stable MaVaN scenarios \cite{Bja08}.
However, due to its somewhat complex analytical description, the GCG fluid is not the most suitable for this kind of quantitative comparison.

We observe that the interval of parameters $A_{\s}$, $m_{\0}$ and eventually $\alpha$, for which our approximation can be applied ($\epsilon < 1$), is valid for $a > a_{\N\R}$ and severally constrained by the imposition $a_{\N\R}\lesssim 1$.
For phenomenologically viable values of $A_{\s}$ ($ 0.7 \lesssim A_{\s} \lesssim 1)$  \cite{Ber03,Ber05} one finds that $0.1 < \epsilon \lesssim 1$.
In general terms, just under quite special circumstances the usual SC and the perturbative contribution of MaVaN's match.
In the original MaVaN scenario \cite{Far04}, the SC corresponds to the adiabatic solution ($H^{\2} \ll \mbox{d}^{\2}V/\mbox{d}\phi^{2}$) of the scalar field equation of motion.
In this case, the kinetic energy terms of the scalar field can be safely neglected.
The consistency of our perturbative scenario with the stationary condition can be achieved only when the kinetic energy contribution is not relevant at late times.
From Eqs.~(\ref{gcg21}), (\ref{gcg21B}) one can easily recover the kinetic energy component of the GCG fluid in terms of $V\bb{\phi\bb{a}}$ and $\rho_{\phi}$,
\begin{equation}
\frac{\dot{\phi}^{\2}}{2\, \rho_{\phi}} = 1 - \frac{V\bb{\phi\bb{a}}}{\rho_{\phi}} =
\frac{1}{2}\left[1 + \frac{a^{\3(\1+\alpha)}\,A_{\s}}{1-A_{\s}}\right]^{\mi \1},
\label{gcg21F}
\end{equation}
which can be set equal to $(1 - A_{\s})/2$ at present times ($a = 1$).
Turning back to Eq.~(\ref{gcg21}), one notices that, for $A_{\s} =0$, the GCG behaves always as matter, whereas for $A_{\s} =1$, it behaves always as a cosmological constant.
Consequently, it is natural that the relevance of the kinetic energy term at present times is suppressed when the parameter $A_{\s}$ gets close to unity, which enlarges the agreement between our perturbative approach and the SC analysis.

\begin{figure}
\vspace{-0.5 cm}
\centerline{\psfig{file=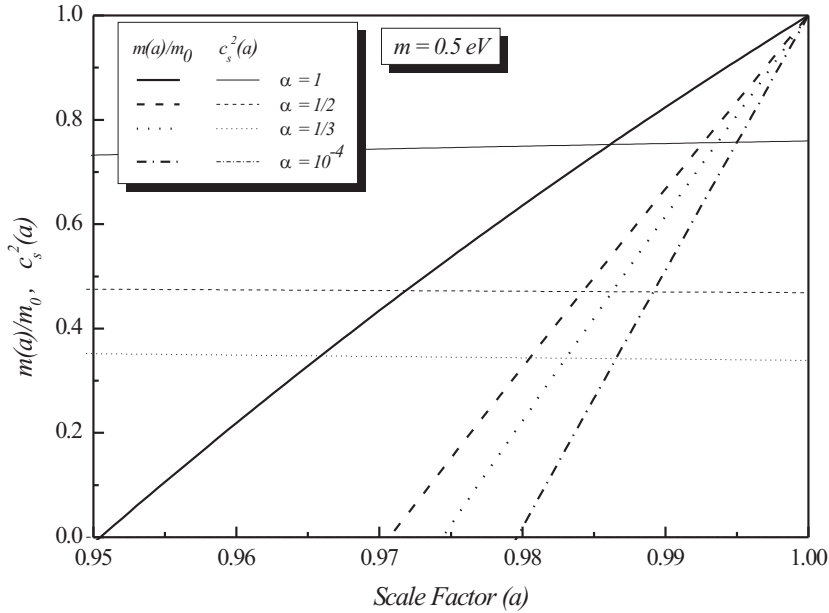,width=14cm}}
\vspace{-0.5 cm}
\caption{Model independent perturbative modification on squared speed of sound $c^{\2}_{s}$ as a function of the scale factor for the neutrino-GCG coupled fluid in comparison with the adiabatic GCG fluid for $A_{\s} = 0.7$ with $\alpha = 1, 1/2,\,1/3,\, 10^{\mi\4}$, for a present-day value of neutrino mass, $m_{\0} = 0.5\, eV$.}
\label{fGCG-10}
\end{figure}

In what concerns the results of our approach, we illustrate the increasing the neutrino mass with the scale factor for a set of phenomenologically consistent parameters for the GCG in Fig.~\ref{fGCG-10}.
Interestingly, from Fig.~\ref{fGCG-10}, for $m_{\0} = 0.5\, eV$, a fairly typical value, we can see that stable MaVaN perturbations correspond to a well defined effective squared speed of sound,
\begin{equation}
c_{s}^{\2} \simeq \frac{d p_{\phi}}{d(\rho_{\phi}+ \rho_{\nu})} > 0.
\label{gcg33B}
\end{equation}
The greater are the $m_{\0}$ values, the more important are the corrections to the squared speed of sound, up to the limit where the perturbative approach breaks down.
However, we find that as far as the perturbative approach is concerned, our model does not run into stability problems in the NR neutrino regime.
In opposition, in the SC treatment, where neutrinos are just coupled to dark energy, cosmic expansion in combination with gravitational drag due to cold dark matter have a major impact on the stability of MaVaN models.
Usually, for a general fluid for which the equation of state is known, the dominant behaviour on $c_{s}^{\2}$ arises from the dark sector component and not by the neutrino component.
For the models where the SC (cf. Eq.~(\ref{gcg07})) implies in a cosmological constant-type equation of state, $ p_{\phi} = - \rho_{\phi}$, one obtains $c_{s}^{\2} = -1$ from the very beginning of the analysis.

Moreover, our perturbative approach is in agreement with the assumption that the coupling between neutrinos and dark energy (and/or dark matter) is weak.
We have observed that the stability condition related to the squared speed of sound of the coupled fluid is predominantly governed by dark energy equation of state.
Such a troublesome behaviour should in fact have already been observed in the context of the FNW scenario as the SC implies that $c_{s}^{\2} = -1$ from the very start and the role of recovering the $c^2_{s (\nu + \phi)} > 0$ condition is relegated to the neutrino contribution \cite{Tak06}.
The loosening of the stationary constraint Eq. (\ref{gcg07}) emerges from the dynamical dependence on $\varphi$, more concretely due to a kinetic energy component \cite{Ber08b}.
The knowledge of the background fluid equation of state for the dark sector (the GCG in our example), and the criterion for the applicability of the perturbative approach, do allow us to overcome the $c^2_{s}$ negative problem, independently from the neutrino mass dependence set by the SC.

As a general final remark, we mention that coupling a dark sector field (dark energy or dark energy plus dark matter) to SM states may bring important insights about the physics beyond the SM.
Neutrino cosmology, in particular, is a fascinating example from where salient questions concerning SM particle phenomenology can be resolved or, at least, better understood if one considers the coupling to the dark sector scalar field.

\begin{acknowledgments}
A. E. B. thanks the financial support from the FAPESP (Brazilian Agency) grant 07/53108-2 and the hospitality of the Physics Department of the Instituto Superior T\'{e}cnico, Lisboa, Portugal.
O. B. would like to acknowledge the partial support of Funda\c{c}\~ao para Ci\^encia e Tecnologia (Portuguese Agency)
under the project POCI/FIS/56093/2004.
\end{acknowledgments}

\end{document}